\documentstyle[12pt]{article}

\newcommand{\beq}{\begin{equation}}
\newcommand{\eeq}{\end{equation}}
\newcommand{\lab}{\protect\label}
\newcommand{\cutoff}{\Lambda_\chi}
\newcommand{\capt}{\protect\caption}

\newcommand{\de}{\stackrel{\rightharpoonup}{d}}

\begin{document}

\begin{titlepage}
\begin{flushright}
LNF 94/031 (P)\\
ROM2F 94/12\\
hep-ph/9408231\\
\end{flushright}
\vspace{0.5cm}
\begin{center}

{
\bf
The hadronic vacuum polarization contribution to the \\
muon g-2 in the quark-resonance model}

\vspace*{1.0cm}

\vspace*{1.2cm}
         {\bf E. Pallante\footnotemark
      \footnotetext{ email: pallante@vaxtov.roma2.infn.it \\
                     \hspace*{6.5mm}fax: +39-6-9403-427 }} \\
{\footnotesize{ I.N.F.N., Laboratori Nazionali di Frascati,
Via E. Fermi, 00044 Frascati ITALY.}}

\end{center}

\vspace*{1.5cm}

\begin{abstract}

The hadronic vacuum polarization contribution to the anomalous magnetic
moment of the muon is parametrized by using the quark-resonance model
formulated in \cite{QR}.  In this context a recent prediction obtained
within the ENJL model \cite{RAF} can be affected by two additional
contributions: the next to leading corrections in the inverse cutoff
expansion and the gluonic corrections. Motivated by the necessity of
reaching a highly accurate theoretical prediction of the hadronic
contribution to the muon g-2, we study in detail both the effects.

\end{abstract}
\vspace*{1.0cm}
\hspace*{1.0cm} Short Title: The muon g-2 in the Quark-Resonance model.
\vspace*{2.0cm}
\begin{flushleft}
LNF 94/031 (P)\\
ROM2F 94/12\\
\end{flushleft}
\footnotetext{
Work partially supported by the EEC Human Capital and Mobility program.}
\end{titlepage}
\newpage

Effective chiral Lagrangians {\em \`a la} Nambu-Jona Lasinio are a good
theoretical framework to understand the hadronic vacuum polarization
contribution to the anomalous magnetic moment of the muon, given by the
diagram in Figure 1.

Phenomenological estimates of $a^h_\mu =(g^h-2)/2$ are obtained from the
best fit of the $e^+e^-\to hadrons$ total cross section $\sigma^h(t)$
through the usual dispersion relation

\beq
a_\mu^h = {1\over 4\pi^3 }\int_{4m_\pi^2}^\infty~dt~K(t)~\sigma^h(t),
\eeq

whith the QED function $K(t)$ given by:

\beq
K(t) = \int_0^1~dx~{x^2(1-x)\over x^2+(1-x){t/ m_\mu^2}} .
\eeq

The most recent numerical estimates give the following values:

\begin{eqnarray} && \bullet ~~7.07(.066)(.17)\cdot 10^{-8}~~~~~~
\cite{Kinoshita} \nonumber\\ &&\bullet ~~6.84(.11)\cdot 10^{-8}
{}~~~~~~~~~~~~~~\cite{Casas} \nonumber\\ &&\bullet ~~7.100(.105)(.49)\cdot
10^{-8}~~~~ \cite{Kurdadze} , \lab{PHEN} \end{eqnarray}

where the first error is statistical and the second one is systematic.
In what follows we give a theoretical picture to understand the above
numerical values.

$a^h_\mu$ is related to the renormalized hadronic photon self-energy
$\Pi^h_R(Q^2)$ through the following integral \cite{RAF,RAFOLD}

\beq
a^h_\mu = {\alpha\over \pi}\int_0^1~dx (1-x) \biggl [ -e^2~\Pi^h_R\biggl (
{x^2\over 1-x} m_\mu^2\biggr )\biggr ] .
\lab{INTE}
\eeq

$\Pi^h_R(Q^2)$ is given in terms of the vector two point function
$\Pi_V^1(Q^2)$ which has been extensively analyzed in refs.
\cite{QR,2point}:

\beq \Pi^h_R(Q^2) = \sum_{i=u,d,s}Q_i^2~ (\Pi_V^1(Q^2)-\Pi_V^1(0))=
{2\over 3} (\Pi_V^1(Q^2)-\Pi_V^1(0)) , \eeq

where $Q_i=(2/3,-1/3,-1/3)$ are the charges of the SU(3) flavour quarks
u,d,s and the renormalized photon self-energy satisfies the constraint
$\Pi^h_R(0)=0$.

Because the typical momenta of the off-shell photons are of the order of
the squared muon mass $(Q^2\sim .01 GeV^2)$ the integral is dominated by
the low energy contribution to $\Pi^h_R(Q^2)$.  By doing the Taylor
expansion of $\Pi^h_R(Q^2)$ at $Q^2=0$ and by imposing $\Pi^h_R(0)=0$
one has

\beq \Pi^h_R(Q^2) = Q^2\Pi_R^{h\prime} (Q^2)=Q^2 \biggl [ {d\Pi^h_R\over
dQ^2}(0) + {1\over 2}Q^2{d^2\Pi^h_R\over (dQ^2)^2}(0) +~ ....~\biggr ] .
\eeq

Notice that $\Pi_R^{h \prime} (Q^2)$ coincides with the first derivative
only at $Q^2=0$. The term with the second derivative contains one
additive power of the ratios $Q^2/\cutoff^2$ and $Q^2/M_Q^2$, where
$\cutoff\sim 1$ GeV is the ultraviolet low-energy cutoff and $M_Q\sim
300$ MeV is the infrared low-energy cutoff and are the natural
dimensionful parameters of the long-distance expansion.

By first approximation $\Pi^h_R(Q^2)\sim Q^2 {d\Pi^h_R\over dQ^2}(0)$
and the integral of eq. (\ref{INTE}) gives the value for $a^h_\mu$

\beq
a^h_\mu = \biggl ( {\alpha\over \pi}\biggr )^2 m_\mu^2 ~{4\pi^2\over 3}
\biggl [ -{2\over 3}{d\Pi_V^1\over dQ^2}(0)\biggr ] .
\lab{FIRSTA}
\eeq

The calculation of $a_\mu^h$ reduces then to the calculation of the
first derivative of the vector two-point function at $Q^2=0$, as was
already pointed out in ref. \cite{RAF}.  In ref. \cite{2point} the
long-distance behaviour of the vector function was derived in the ENJL
framework.  In \cite{QR} we have shown that the inclusion in the
bosonized NJL Lagrangian of higher dimensional quark-resonance vertices,
which are suppressed respect to the leading lowest dimensional
four-quark operator by powers of the inverse cutoff $\cutoff$,
 generates next-to-leading power corrections to the leading logarithms
(NPLL) of the parameters of the effective resonance Lagrangian which are
responsible of their $Q^2$ dependence.

As a consequence, higher dimensional quark-resonance operators modify
the long distance behaviour of the vector Green's function predicted by
the ENJL model and enter in the determination of $a_\mu^h$ through eq.
(\ref{INTE}).  Already in the first approximation of eq. (\ref{FIRSTA})
the NPLL corrections proportional to $Q^2$, i.e. of the type
$Q^2/\cutoff^2\ln (\cutoff^2/M_Q^2 )$ give contribution to the
derivative at $Q^2=0$; the derivative at $Q^2=0$ is sensitive to the
behaviour at shorter distances.

We proceed as follows. Using the first approximation of eq.
(\ref{FIRSTA}) we review the ENJL prediction already derived in
\cite{RAF}. Then we study the relevance of two sources of corrections:
the NPLL contributions and the gluon contributions.  Beyond the first
approximation we analyze the sensitivity to both of them via the
evaluation of the dispersive integral of eq. (\ref{INTE}) over the
long-distance part of $\Pi^h_R(Q^2)$.

The vector two-point function in the ENJL model and in the chiral limit
can be parametrized in terms of the running photon-vector coupling
$f_V(Q^2)$ and the squared mass of the vector resonance $M_V^2(Q^2)$ as
follows:

\beq
\Pi_V^1(Q^2) = {2f_V^2(Q^2)M_V^2(Q^2)\over M_V^2(Q^2)+ Q^2},
\lab{VEC}
\eeq

or equivalently in terms of the vector two-point function in the mean
field approximation $\overline{\Pi}_V^1(Q^2)$ \cite{2point}

\beq
\Pi_V^1(Q^2) ={ \overline{\Pi}_V^1(Q^2)\over 1+ Q^2
{8\pi^2 G_V\over N_c\cutoff^2} \overline{\Pi}_V^1(Q^2) } ,
\lab{MF}
\eeq

where $G_V(\cutoff )$ is the coefficient of the four-quark vector-like
interaction in the ENJL Lagrangian.  The $Q^2$ dependent parameters are
given by:

\begin{eqnarray} \overline{\Pi}_V^1(Q^2)&=& {N_c\over 16\pi^2} 8
\int_0^1~d\alpha~\alpha (1-\alpha )~ \Gamma (0, \alpha_Q) \nonumber\\
f^2_V(Q^2)&=&{1\over 2}\overline{\Pi}_V^1(Q^2) \nonumber\\ M_V^2(Q^2)&=&
{N_c\cutoff^2\over 8\pi^2 G_V} \biggl (\overline{\Pi}_V^1(Q^2)\biggr
)^{-1} , \lab{PARA} \end{eqnarray}

with the incomplete Gamma function $\Gamma (0,\alpha_Q) = -\ln\alpha_Q
-\gamma_E + {\cal{O}}(\alpha_Q)$, $\alpha_Q = (M_Q^2 + \alpha (1-\alpha
)Q^2)/ \cutoff^2$ and $\gamma_E = 0.5772...$ is the Euler's constant.
Expressions (\ref{VEC}) and (\ref{MF}) with the parameters (\ref{PARA})
correspond to the diagram of Figure 2a, which is the infinite
resummation of linear chains of constituent quark bubbles with the
insertion of the leading four-quark vector operator of the ENJL
Lagrangian.  Linear chains of quark bubbles are of order $N_c$, while
loops of chains of quark bubbles are of order 1 in the $1/N_c$
expansion.

The derivative at $Q^2=0$ is given by

\beq {d\Pi_V^1\over dQ^2}(0)={d\overline{\Pi}_V^1\over dQ^2}(0)-
{\overline{\Pi}_V^1}(0)^2 {8\pi^2 G_V\over N_c\cutoff^2 }
 = 2{df_V^2\over dQ^2}(0) - 2 {f_V^2(0)\over M_V^2(0)} .
\lab{DEFI}
\eeq

The ENJL model gives the following prediction:

\beq {d\Pi_V^1\over dQ^2}(0) = -{N_c\over 16\pi^2}{1\over M_Q^2} {4\over
15} \biggl [ e^{-{M_Q^2\over \cutoff^2}} + {5\over 6} {1-g_A\over g_A}
\Gamma (0, {M_Q^2\over \cutoff^2})\biggr ] , \lab{NJLP} \eeq

where $g_A$ is the mixing parameter between the axial-vector and the
pseudoscalar mesons in the bosonized ENJL action and it is related to
the vector meson mass and the vector coupling $G_V$ by the following
relations \cite{2point}:

\beq {1-g_A\over g_A}= 4M_Q^2{G_V\over\cutoff^2}\Gamma
(0,M_Q^2/\cutoff^2 ) ={6M_Q^2\over M_V^2}.  \lab{RELA} \eeq

With the values of the best fit 1 of ref. \cite{ENJL} $M_Q=265 MeV$,
$\cutoff = 1.165 GeV$ and $g_A = 0.61$ we obtain for $a_\mu^h$ the value
$a^h_\mu = 8.66\cdot 10^{-8}$, which corresponds to the value
${d\Pi_V^1\over dQ^2}(0)=-0.164$ of the first derivative.  In formula
(\ref{NJLP}) the incomplete Gamma function has been approximated to its
leading logarithmically divergent part.  This corresponds to the
substitutions $\Gamma (0,\alpha_Q) = -\ln\alpha_Q -\gamma_E$ and
$exp(-M_Q^2/ \cutoff^2)=1$, where $exp(-M_Q^2/ \cutoff^2)$ comes from
the derivative of $\Gamma (0,\alpha_Q)$ at $Q^2=0$.

Chiral loop corrections due to $\pi , K$ exchanges, are not included in
the above value of $a_\mu^h$. They are given by the diagram in Figure 2b
and are next-to-leading (i.e. O(1)) in the $1/N_c$ expansion; in the
ENJL framework they are generated by diagrams with loops of chains of
quark bubbles. This contribution has been derived in \cite{RAF} using
{\em ChPt} with a value of $(0.71\pm 0.07)\cdot 10^{-8}$.  The two
summed contributions give

\beq a_\mu(had) = a_\mu ({\hbox{Fig. 2a}}) + a_\mu(\chi {\hbox{loops}})
= 9.37\cdot 10^{-8}, \lab{NUM1} \eeq

which is a rather high value compared to the phenomenological estimates
in (\ref{PHEN}). Beyond the fact that the first derivative approximation
can be not sufficiently accurate, we first analyze the effects of the
two ``extra'' contributions we have mentioned.

The first source of ``extra'' corrections can be derived in the
framework of the Quark-Resonance model formulated in \cite{QR}.

The vector two-point function calculated with the inclusion of NTL
vertices corresponds to the diagram of Figure 2c.: the vector resonance
exchange diagram plus the ``local'' diagram. The infinite resummation of
linear chains of quark bubbles of Figure 2a is a part of the
contributions to the renormalized vector resonance propagator.  
the insertion 
vector four-quark 

The complete set of NTL corrections (i.e. $1/\cutoff^2$), includes two
types of contributions \cite{QR}:

\begin{description} \item[I.] NTL genuine power corrections (NTLP) of
the type: ${Q^2\over \cutoff^2 },{M_Q^2\over \cutoff^2 }$

\item[II.] NTL power corrections to the leading logs (NPLL) of the type:
 ${Q^2\over \cutoff^2 }\ln \alpha_Q , {M_Q^2\over \cutoff^2 }\ln
\alpha_Q$.  \end{description}

Class I is generated by an infinite number of higher dimensional
quark-resonance vertices, while class II is generated by a finite number
of $1/\cutoff^2$ vertices \cite{QR}.
 A best fit to the experimental $e^+e^-\to hadrons$ cross section in the
$I=1,J=1$ channel and in the intermediate $Q^2$ region (500 MeV - 900
MeV) has shown the numerical relevance of the new NPLL counterterms
proportional to $Q^2$ \cite{QR}, while NPLL corrections proportional to
the IR cutoff $M_Q$ are negligible in this regime. They are not
negligible when treating the very low energy behaviour $(Q^2\to 0)$ of
the Green's functions, as it happens in the present case.

In table (\ref{TERMS}) the type of quark-resonance vertices we need are
listed.  We distinguish four sectors: the derivative one, the vector
one, the scalar one and the scalar-vector one.  NTL corrections
proportional to $Q^2$ get contributions from the derivative and the
vector sets, while NTL corrections proportional to $M_Q^2$ get
contributions also from the scalar and scalar-vector sets when the
scalar field assumes its VEV, $<H>=M_Q$, which plays the role of the IR
cutoff of the effective theory.

The full running of the vector resonance parameters ($Z_V =$ vector wave
function renormalization constant, $f_V$ = vector-photon coupling and
$M_V$ = vector mass) is parametrized as follows:

\begin{eqnarray} Z_V(Q^2)&=& {N_c\over 16\pi^2} {1\over 3}
\int_0^1~d\alpha~ \Gamma (0,\alpha_Q ) \biggl [ 6\alpha (1-\alpha ) +
12\alpha (1-\alpha )\beta_V^1 {Q^2\over \cutoff^2} + \beta_M^1
{M_Q^2\over \cutoff^2} P_1(\alpha )\biggr ] \nonumber\\
M_V^2(Q^2)&=&{N_c\over 16\pi^2}{\cutoff^2\over 2{\tilde{G}_V}}{1\over
Z_V} \biggl [ 1 + \beta_M^2 {M_Q^2\over \cutoff^2} \int_0^1~d\alpha~
\Gamma (0,\alpha_Q ) P_2(\alpha )\biggr ] \nonumber\\ f_V(Q^2)&=&
{1\over \sqrt{Z_V}} {N_c\over 16\pi^2}{\sqrt{2}\over 3}
\int_0^1~d\alpha~ \Gamma (0,\alpha_Q ) \biggl [ 6\alpha (1-\alpha ) +
6\alpha (1-\alpha )\beta_\Gamma^1
 {Q^2\over \cutoff^2} +\nonumber\\
&& 6\alpha (1-\alpha )\beta_V^1
{Q^2\over \cutoff^2} +\beta_M^3 {M_Q^2\over \cutoff^2} P_3 (\alpha
)\biggr ] .  \lab{QRPAR} \end{eqnarray}

In the calculation of $f_V$ and $Z_V$ we have not included NTLP
corrections.  They give a correction which is almost $1\%$ of the
leading logarithmic term.  Genuine power corrections can be generated by
the quadratically or more divergent part of diagrams that contain higher
dimensional vertices which are suppressed by inverse powers of the UV
cutoff $\cutoff$.

NTLP corrections to the squared vector mass and proportional to $M_Q$
can be reabsorbed in the renormalization of the vector coupling $G_V$ as
follows:

\beq {\cutoff^2\over \tilde{G}_V} = {\cutoff^2\over G_V} \biggl [ 1+
\delta^\prime {M_Q^2\over \cutoff^2}\biggr ] .
\eeq

In eq. (\ref{QRPAR}) the dependence upon the Feynman parameter $\alpha$
is always of the form $\alpha (1-\alpha )$ for the $Q^2/\cutoff^2$
corrections.  The polynomials $P_i(\alpha)$ are explicitely calculable
for each NPLL counterterm.  We will put them equal to the leading
polynomial $\alpha (1-\alpha)$ which is still a good approximation. With
this assumption the three vector parameters $Z_V$, $M_V^2$ and $f_V^2$
at $Q^2=0$ are given by:

\begin{eqnarray} Z_V(0)&=& {N_c\over 16\pi^2} {1\over 3} \Gamma
(0,M_Q^2/\cutoff^2 ) \biggl [ 1 +{\beta_M^1\over 6} {M_Q^2\over
\cutoff^2} \biggr ] \nonumber\\
M_V^2(0)&=&{N_c\over 16\pi^2}
{\cutoff^2\over 2{\tilde{G}_V}} {1\over Z_V(0)} \biggl [ 1+
{\beta_M^2\over 6} {M_Q^2\over \cutoff^2} \Gamma (0,M_Q^2/\cutoff^2 )
\biggr ]
   \nonumber\\
f_V^2(0)&=&{2\over 9}{1\over Z_V(0)} \biggl ({N_c\over
16\pi^2}\biggr )^2 \Gamma^2 (0, M_Q^2/\cutoff^2 )\biggl [ 1+{
\beta_M^3\over 6} {M_Q^2\over \cutoff^2} \biggr ]^2 .  \end{eqnarray}

The NPLL corrections proportional to $M_Q^2$ do modify the ENJL leading
prediction of the parameters of the effective meson Lagrangian at zero
energy.  One way to estimate the new coefficients $\beta_M^i$, is to do
the best fit of the whole set of the leading $(M_Q, \cutoff ,g_A)$ and
NPLL parameters (with and without gluonic corrections) at $Q^2=0$ using
as inputs the experimental values of the low energy meson parameters;
although a sufficiently accurate determination of the $\beta_M^i$ is
still not accessible with the present uncertainty on the very low energy
experimental data.

The ratio $f_V^2/M_V^2$ and the derivative of the squared coupling at
$Q^2=0$, which enter in eq. (\ref{DEFI}), are given by the following
expressions and retaining up to $1/\cutoff^2$ terms:

\begin{eqnarray}
{f_V^2(0)\over M_V^2(0)}&=&{2\over 3} {N_c\over 16\pi^2} {1\over
{M_V^2(0)}^{ENJL}} \Gamma (0, M_Q^2/\cutoff^2 )\biggl [ 1+
 { \beta_M^3\over 3}{M_Q^2\over \cutoff^2} -{\beta_M^2\over
6}{M_Q^2\over \cutoff^2}\Gamma (0, M_Q^2/\cutoff^2 ) \biggr ]
\nonumber\\ f_V^{2\prime}(0)&=&{2\over 3}{N_c\over 16\pi^2}\biggl \{
\Gamma (0, M_Q^2/\cutoff^2 ){2\beta_\Gamma^1\over \cutoff^2} -{1\over
5M_Q^2} e^{-{M_Q^2/\cutoff^2}}
 \biggl [ 1+ {2\beta_M^3 -\beta_M^1\over 6}{M_Q^2\over \cutoff^2} \biggr
]\biggr \} .
\lab{23}
\end{eqnarray}

Again the approximation $\Gamma (0,\alpha_Q) = -\ln\alpha_Q -\gamma_E$
and $exp(-M_Q^2/ \cutoff^2)=1$ is understood and we have used the
relation of eq. (\ref{RELA}) which defines ${M_V^2(0)}^{ENJL}$. We
assume its numerical value given by the fit 1 of ref.  \cite{ENJL} which
is $(0.811)^2~GeV^2$.

The size of the $Q^2$ corrections has been determined by the best fit of
ref. \cite{QR}:

\beq
\beta_\Gamma^1 = -0.75\pm 0.01~~~~~~~~~~~\beta_V^1=-0.79\pm 0.01.
\eeq

The sign of $\beta_\Gamma^1, \beta_V^1$ is negative and increases the
value of $a^h_\mu$. By including only $Q^2$ type corrections the value
of $a^h_\mu$ increases to $1.22\cdot 10^{-7}$.  Comparable values of the
$\beta_M^i$ coefficients give as a maximum range of variation of
$a_\mu^h$ $1.20\cdot 10^{-7} \div 1.24\cdot 10^{-7}$. This proves that
$Q^2$ dependent contributions give the bulk of the NPLL corrections to
$a_\mu^h$ calculated within the first derivative approximation.

Gluonic corrections can be parametrized following ref. \cite{2point}.
The leading contribution in the $1/N_c$ expansion involves only one
unknown parameter $g$ which is related to the lowest dimensional gluon
vacuum condensate:

\beq
g = {\pi^2\over 6N_c M_Q^4} <{\alpha_s\over \pi} GG> .
\lab{GLU}
\eeq

Because $<{\alpha_s\over \pi} GG>$ is $O(N_c)$ $g$ is $O(1)$ and the
leading gluonic correction to the two-point vector function is still
$O(N_c)$.  The role of gluonic corrections in effective low energy
fermion models is still an unsolved theoretical problem. What is clear
is that the two gluon condensate of eq. (\ref{GLU}) is only the low
energy ($<\cutoff$) `residue' of the standard two gluon condensate which
is phenomenologically estimated through QCD sum rules. The latter
suggest an $O(1)$ $g$ parameter, while best fits of the ENJL parameters,
using as inputs the experimental values of $f_\pi , L_i$ and the meson
resonances' parameters \cite{ENJL}, strongly favour a value of $g\le
0.5$.  The first phenomenological estimation, usually referred to as the
``standard value'', has been obtained by Shifman et al. (SVZ) \cite{SVZ}
by studying the charmonium channel and using Operator Product Expansion:
they obtain $<{\alpha_s\over \pi} GG> = 0.012~GeV^4$ (i.e. g=1.3).  A
compatible value $<{\alpha_s} GG>= (3.9\pm 1.0) 10^{-2}~GeV^4$ has been
obtained \cite{Launer} from the $e^+e^-\to I=1$ hadron cross section and
using moment sum rules ratio.  The most recent estimation via FESR
(Finite Energy sum rules) \cite{FESR} gives a significative higher value
$<{\alpha_s\over \pi} GG> = 0.044^{+ 0.015}_{-0.021}~GeV^4$ (i.e. $g\sim
4.8$), although the error in all cases has to be conservatively taken
around $40\%$.

In what follows we will use three values of g which are acceptable in
the ENJL model. The IR cutoff $M_Q$, the UV cutoff $\cutoff$ and the
axial-pseudoscalar mixing parameter $g_A$, extracted from the best fits
of experimental data, vary as functions of g as it is summarized in
table (\ref{GDIP}).  $M_Q$ decreases sensitively by increasing g.

Leading gluonic corrections can be expressed as an additive contribution
to the function $\overline{\Pi}_V^1(Q^2)$ which is written for $N_c=3$
as \cite{Narison}

\beq
\overline{\Pi}_V^{1g}(Q^2)= {3gM_Q^4\over 2\pi^2 Q^4} (-1 + 3 {\cal
I}_2 -2{\cal I}_3 ),
\eeq

\noindent in terms of the $Q^2$ dependent functions

\beq
{\cal I}_N = \int_0^1~d\alpha~
{1\over [ 1+ {Q^2\over M_Q^2} \alpha (1-\alpha )]^N} .
\eeq

\noindent Expanding at small $Q^2$ we obtain their values at $Q^2=0$:

\beq
\overline{\Pi}_V^{1g}(0)= -{3g\over 20\pi^2}
{}~~~~~~~~~\overline{\Pi}_V^{\prime 1g}(0)= {6g\over 70\pi^2M_Q^2} ,
\eeq

\noindent which give the following expression for the derivative:

\begin{eqnarray}
{d\Pi_V^1\over dQ^2}(0) &=& {d{\Pi_V^1}^ {g=0}\over
dQ^2}(0) +\Delta^g \nonumber\\
\Delta^g&=& {6g\over 70\pi^2M_Q^2} -
\biggl ({3g\over 20\pi^2}\biggr )^2 {8\pi^2 G_V\over N_c\cutoff^2}
\biggl ( 1- {40\pi^2\over 3g} {\Pi_V^1}^ {g=0}(0)\biggr ) .
\lab{DGLUON}
\end{eqnarray}

Notice that ${\Pi_V^1}^{g=0}(0) = {\overline{\Pi}_V^1}^{ g=0}(0)$; all
the quantities with superscript $g=0$ are those at g=0 with the
parameters $M_Q$, $\cutoff$ and $g_A$ rescaled according to table
(\ref{GDIP}) for a given g in formula (\ref{DGLUON}).  In table
(\ref{GDERT}) we give the gluonic corrections to the derivative at
$Q^2=0$ with g = 0.25, 0.5.

Gluonic contributions decrease $a_\mu^h$ towards a better agreement with
the phenomenological estimates (\ref{PHEN}).

The evaluation of the full dispersive integral (\ref{INTE}) requires the
knowledge of the long distance (ld) plus the short distance (sd)
behaviour of $\Pi_R^h(Q^2)$. We can do our best performing the matching
between the ld prediction coming from the effective theory and the sd
prediction coming from perturbative QCD.  In ref. \cite{RAF} a value of
$a_\mu^h(Fig. 2a)=6.7\cdot 10^{-8}$ has been obtained in the ENJL
framework with a best fitted matching point $\hat{x}\simeq 0.91$ which
corresponds to an euclidean
$\hat{Q}^2=\hat{x}^2/(1-\hat{x})m_\mu^2\simeq (320)^2~MeV^2$.  The value
obtained for $a_\mu^h$ is quite better than the first approximation
value $8.66\cdot 10^{-8}$.

To see how ``extra'' corrections modify the ENJL prediction is
sufficient to study the integral (\ref{INTE}) over the ld part in the
range $\int_0^{\hat{x}}$ for different values of $\hat{x}$ corresponding
to an equivalent value of $\hat{Q}^2 = (0.3)^2,(0.5)^2, (0.8)^2~GeV^2$.

NPLL corrections proportional to $Q^2$ in the QR model lead to a
 vector two-point function which decreases faster in $Q^2$ then the ENJL
prediction.  The dispersive integral gives for $a_\mu^h$ the values of
table (\ref{QRTOT}).

The correction induced by higher order terms proportional to $Q^2$ and
which are relevant in the intermediate $Q^2$ region is about one
percent.  This proves that $a_\mu^h$ is practically only sensitive to
perturbative corrections which modify the very low $Q^2$ region, i.e.
$Q^2\leq (500MeV)^2$.

Gluonic corrections modify the ENJL vector two-point function for all
$Q^2$ as shown in figure 3. They modify the ENJL prediction of $a_\mu^h$
as summarized in table (\ref{GTOT}).  The corrections are $10\%$ for
g=0.5 and $15\%$ for g=0.25. They decrease $a_\mu^h$. A better
determination of non-gluonic contributions to $a_\mu^h$ can be used to
constrain the value of the g parameter in low energy effective fermion
models.

We conclude that the ENJL prediction is reliable within $30\%$.  Both
gluonic corrections and next-to-leading higher dimensional
quark-resonance interactions have to be taken into account if one wants
to reach an accuracy better than $30\%$.  The first derivative
approximation is not sufficiently accurate to estimate the hadronic
vacuum polarization contribution to $a_\mu^h$, although the quantity is
sensitive only to ``extra'' corrections in the low $Q^2$ region ($Q <
500$ MeV).  NPLL corrections proportional to $Q^2$ do not affect
$a_\mu^h$.  NPLL corrections proportional to $M_Q^2$ and gluonic
corrections are relevant in the low $Q^2$ region.  Their inclusion can
explain the full agreement of the ENJL prediction with the
phenomenological estimates (\ref{PHEN}) of $a_\mu^h$.

\vspace{3cm}
{\bf Acknowledgements}. I am grateful to Eduardo de Rafael for having
called my attention to this problem and for useful discussions.
\vfill\eject

\centerline {{\bf FIGURE CAPTIONS}}

\begin{description}

\item[1)] Hadronic vacuum polarization diagram to the anomalous
magnetic moment of the muon.

\item[2a)] The vector two-point function in the ENJL model and leading
in the $1/N_c$ expansion. It is given by the infinite resummation of
linear chains of constituent quark bubbles; each four-quark vertex is
the leading vertex of the ENJL model with coupling $G_V$.

\item[2b)] Chiral loops corrections to the vector two-point function.
They are of O(1) (i.e. next-to-leading) in the $1/N_c$ expansion and in
the ENJL model they are generated by the infinite resummation of loops
of chains of constituent quark bubbles.

\item[2c)] The vector two-point function in the QR model and leading in
the $1/N_c$ expansion. It is given by the ``local'' diagram and the
vector-exchange diagram with the renormalized vector meson propagator.
 The diagram of figure 2a is a part of the contribution to the renormalized
vector meson propagator.

\item[3)] The vector two-point function $\Pi_V^1(Q^2)$ in the ENJL model
without gluonic corrections (solid line) and with gluonic corrections
for g=0.25, 0.5 (dashed lines).

\end{description}

\vskip 2.truecm

\centerline {{\bf TABLE CAPTIONS}}

\begin{description}

\item[1)] $1/\cutoff^2$ quark-resonance vertices of the sectors: I)
derivative, II) vector, III) scalar and IV) scalar-vector which give
contribution to the NPLL corrections.

\item[2)] Values of $M_Q$, $\cutoff$ and $g_A$ of the ENJL model
obtained from the best fits of ref. \cite{ENJL} to the experimental data
of low energy parameters and with fixed values of the gluonic paramater
g=0, 0.25, 0.5.

\item[3)] Gluonic corrections to $a_\mu^h$ in the first derivative
approximation for two values of the gluonic parameter $g$ favoured by
the ENJL model.  $\Delta^g$ is the gluonic correction to the first
derivative as defined in eq. (\ref{DGLUON}), $d\Pi_V^1/dQ^2(0)$ is the
first derivative including gluonic corrections. In the last two columns
the numerical value of $a_\mu^h$ and the variation in percentage are
shown.

\item[4)] Numerical values of $a_\mu^h$ obtained through the dispersion
relation (\ref{INTE}), where the integral is performed on the
long-distance part of $\Pi_R^h$ predicted by the QR model in the range
$0<Q^2<\hat{Q}^2$ and compared to the ENJL prediction.

\item[5)] Numerical values of $a_\mu^h$ obtained through the dispersion
relation (\ref{INTE}), where the integral is performed on the
long-distance part of $\Pi_R^h$ predicted by the ENJL model in the range
$0<Q^2<\hat{Q}^2$ without the inclusion of gluonic corrections (g=0) and
including gluonic corrections with g=0.25, 0.5.

\end{description}
\vfill\eject

\vfill
\newpage

\begin{table}
\begin{tabular}{l}
DERIVATIVE \\
\\
$\beta_\Gamma^1~\bar{Q}\gamma_\nu d^\mu \Gamma_{\mu\nu} Q
+\beta_\Gamma^2~\bar{Q}\gamma_\mu \{\de^\mu ,\de^2 \} Q $  \\
\\
VECTOR \\
\\
$\beta_V^1 \bar{Q}\gamma_\mu d^2 W_{\mu}^+ Q+
\beta_V^2\bar{Q} \gamma_\mu \{ W_\mu^+ ,\de^2\} Q+
\beta_V^3 \bar{Q}\gamma_\mu \{\{\de_\mu ,\de_\nu\} ,W_{\nu}^+\} Q+
\beta_V^4 \bar{Q}\gamma_\mu  [ \Gamma_{\mu\nu}, W_\nu^+] Q +$ \\
$\beta_V^5 \bar{Q}\gamma_\mu\{ W^{+2}, \de^\mu \} Q +
\beta_V^6 \bar{Q}\gamma_\mu [ d^\mu W^{+}_\nu, W_\nu^+ ]Q+
\beta_V^7 \bar{Q}\gamma_\mu (W^{+}_\mu W^{+}_\nu \de_\nu + \de_\nu
W^{+}_\nu W^{+}_\mu )Q +   $          \\
$\beta_V^8 \bar{Q}\gamma_\mu (W^{+}_\nu W^{+}_\mu \de_\nu + \de_\nu
W^{+}_\mu W^{+}_\nu )Q +
\beta_V^9 \bar{Q}\gamma_\mu [ d^\nu W^{+}_\mu, W_\nu^+ ]Q $\\
            \\
 \\
SCALAR        \\
\\
$\beta_S^1~ \bar{Q}H^3 Q+ \beta_S^2~\bar{Q}\gamma_\mu
\{ H^2, \de^\mu\} Q +
\beta_S^3  ~\bar{Q}\{ H, \de^2 \} Q$ \\
\\
SCALAR-VECTOR            \\
\\
$\beta_{SV}^1 \bar{Q}\{W^{+2},H\} Q
+\beta_{SV}^2 \bar{Q}W_\mu^+ H W_\mu^+ Q
+ \beta_{SV}^3\bar{Q} \gamma_\mu\{ H^2, W_\mu^+\} Q+
\beta_{SV}^4\bar{Q} \gamma_\mu HW_\mu^+ H Q $ \\
$+\beta_{SV}^5 \bar{Q} (W_\mu^+ H \de_\mu + \de_\mu H W_\mu^+)Q
+\beta_{SV}^6 \bar{Q} (HW_\mu^+  \de_\mu + \de_\mu  W_\mu^+H)Q$.
             \\
\end{tabular}
\capt[1]{\lab{TERMS}}
\end{table}
\vskip 3cm

The covariant derivative of the vector field $W_\mu^+$ and the scalar
field $H$ is defined as $d_\mu {\cal O} = \partial_\mu {\cal O}
 + [\Gamma_\mu ,{\cal O}]$, while the covariant derivative on the {\em
constituent} quark field $Q$ is defined as $d_\mu Q = \partial_\mu Q+
\Gamma_\mu Q$.  $\Gamma_\mu = 1/2 (\xi^\dagger \partial_\mu\xi +
\xi\partial_\mu\xi^\dagger )$
 is the vector current constructed with the square root of the
pseudoscalar meson field $U=\xi^2=exp(i\phi/f_\pi )$.  The identity
$\Gamma_{\mu\nu}= -{i\over 2}f_{\mu\nu}^+ +{1\over 4} [\xi_\mu , \xi_\nu
]$ introduces the field strenght of the electromagnetic field
$f_{\mu\nu}^+= \xi F_{\mu\nu}\xi^\dagger +\xi^\dagger F_{\mu\nu}\xi$
which generates the photon-vector interaction associated with the
coupling $f_V$.

\vfill \newpage

\begin{table}
\centering
\begin{tabular}{|c|c|c|c|}\hline
  & & & \\
$g$ & $M_Q$ (MeV)& $\cutoff$ (GeV) & $g_A$\\ \hline
    & & &\\
$g=0$    & 265& 1.165 & 0.61\\
    & && \\
$g=0.25$    & 246& 1.062 & 0.62\\
    & & &\\
$g=0.5$    & 204& 1.090 &0.66\\
    & & \\ \hline
\end{tabular}
\capt[1]{\lab{GDIP}}
\end{table}
\vspace{2cm}

\begin{table}
\centering
\begin{tabular}{|c|c|c|c|c|}\hline
  & & & & \\
$g$ & $\Delta^g$  & ${d\Pi_V^1/dQ^2}(0)$   & $a_\mu^h$  &
$\Delta a_\mu/a_\mu (\% )$ \\ \hline
    & & & & \\
$g=0.25$    & 0.048   & -0.136  &$7.2\cdot 10^{-8}$    & $-17\%$
\\
    & & & & \\
$g=0.5$    & 0.134   &-0.133   & $7.0\cdot 10^{-8}$   &  $-19\%$
\\
    & & & & \\
\hline
\end{tabular}
\capt[1]{\lab{GDERT}}
\end{table}
\vspace{2cm}

\begin{table}
\centering
\begin{tabular}{|c|c|c|}\hline
  & &  \\
$\hat{Q}^2$ $GeV^2$&
$a_\mu^{QR}\times 10^{8}$ & $a_\mu^{ENJL}\times 10^{8}$ \\
\hline
     & &\\
$(0.5)^2$ & 6.9 & 6.8\\
    & &\\
$(0.8)^2$ & 7.3 & 7.1 \\
&  & \\ \hline
\end{tabular}
\capt[1]{\lab{QRTOT}}
\end{table}
\vspace{2cm}

\begin{table}
\centering
\begin{tabular}{|c|c|c|c|}\hline
  & & & \\ $\hat{Q}^2$ $GeV^2$& $a_\mu^h$ g=0 & $a_\mu^h$ g=0.25
&$a_\mu^h$ g=0.5 \\
\hline & & & \\
$(0.3)^2$ & $5.9\cdot 10^{-8}$ &
$5.0\cdot 10^{-8}$ & $5.3\cdot 10^{-8}$ \\ & & & \\ $(0.5)^2$ &
$6.8\cdot 10^{-8}$ & $5.9\cdot 10^{-8}$ & $6.3\cdot 10^{-8}$
\\ & & & \\
$(0.8)^2$ & $7.1\cdot 10^{-8}$ & $6.2\cdot 10^{-8}$ & $6.6\cdot 10^{-8}$
\\ & & & \\ \hline
\end{tabular}
\capt[1]{\lab{GTOT}}
\end{table}

\vfill\eject
\newpage

\newpage

\begin{thebibliography}{99}

\bibitem{QR} E. Pallante and R. Petronzio, Preprint ROM2F 37/93,
submitted to Nucl. Phys.  B

\bibitem{RAF}
E. de Rafael, {\em Phys. Lett. } {\bf B322} (1994) 239.

\bibitem{Kinoshita} T. Kinoshita, B. Ni$\check{z}$i\'c and Y. Okamoto,
{\em Phys. Rev.} {\bf D31}
 (1985) 2108.

\bibitem{Casas} J.A. Casas, C. Lopez and F.J. Yndur\'ain, {\em Phys.
Rev.} {\bf D32} (1985) 736.

\bibitem{Kurdadze}
L.M. Kurdadze et al., {\em Sov. J. Nucl. Phys.} {\bf 40} (1984) 286.

\bibitem{RAFOLD} B.E. Lautrup, A. Peterman and E. de Rafael, {\em Phys.
Rep.} {\bf 3C} (1972) 193.

\bibitem{2point} J. Bijnens, E. de Rafael and H. Zheng, Preprint CERN-TH
6924/93, CTP-93/P2917, NORDITA-93/43 N,P.

\bibitem{ENJL} J. Bijnens, C. Bruno and E. de Rafael, {\em Nucl. Phys.}
{\bf B390} (1993) 501.

\bibitem{SVZ}

M.A. Shifman, A.I. Vainshtein and V.I. Zakharov,
{\em Nucl. Phys.} {\bf B147} (1979) 385, 448.

\bibitem{Launer}

G. Launer, S. Narison and R. Tarrach, {\em Z. Phys.} {\bf C 26} (1984)
433.

\bibitem{FESR}

R.A. Bertlmann et al., {\em Z. Phys.} {\bf C 39} (1988) 231.


\bibitem{Narison}

S. Narison, ``QCD Spectral Sum Rules'', World Scientific Lecture Notes
in Physics, Vol. 26.

\end{thebibliography}
\end{document}